\documentclass[
  journal=pasa,
  manuscript=research-paper, 
  year=202X,
  volume=YY,
]{cup-journal}

\usepackage{amsmath}
\usepackage{amssymb,microtype,siunitx,booktabs}
\usepackage[colorlinks=true,
            citecolor=blue,
            linkcolor=blue,
            urlcolor=cyan,
            anchorcolor=blue]{hyperref}

\usepackage{lipsum}
\usepackage{graphicx}

\title{Untapped: Veloce Detects Calcium in the Atmosphere of WASP-189b}

\author{Nicholas W. Borsato}
\affiliation{School of Mathematical and Physical Sciences, Macquarie University, Sydney, NSW 2109, Australia}
\alsoaffiliation{Astrophysics and Space Technologies Research Centre, Macquarie University, Sydney, NSW 2109, Australia}
\alsoaffiliation{Lund Observatory, Division of Astronomy and Theoretical Physics, Lund University, Box 43, 221 00 Lund, Sweden}
\email[Nicholas W. Borsato]{nicholas.borsato@hdr.mq.edu.au}

\author{Joachim Kr\"uger}
\affiliation{Centre for Astrophysics, University of Southern Queensland, Toowoomba, QLD 4350, Australia}

\author{Daniel B. Zucker}
\affiliation{School of Mathematical and Physical Sciences, Macquarie University, Sydney, NSW 2109, Australia}
\alsoaffiliation{Astrophysics and Space Technologies Research Centre, Macquarie University, Sydney, NSW 2109, Australia}

\author{Simon J. Murphy}
\affiliation{Centre for Astrophysics, University of Southern Queensland, Toowoomba, QLD 4350, Australia}

\author{Duncan Wright}
\affiliation{Centre for Astrophysics, University of Southern Queensland, Toowoomba, QLD 4350, Australia}

\author{Sarah L. Martell}
\affiliation{School of Physics, University of New South Wales, Sydney, NSW 2052, Australia}

\received{15 Aug 2025}
\revised{05 Sep 2025}
\accepted{17 Sep 2025}
\published{}


\begin{document}

\begin{abstract}
High-resolution transmission spectroscopy has become a powerful tool for detecting atomic and ionic species in the atmospheres of ultra-hot Jupiters. In this study, we demonstrate for the first time that the Australian-built Veloce spectrograph on the 3.9-m Anglo-Australian Telescope can resolve atmospheric signatures from transiting exoplanets. We observed a single transit of the ultra-hot Jupiter WASP-189b—a favourable target given its extreme irradiation and bright host star—and applied the cross-correlation technique using standardised templates. We robustly detect ionised calcium (Ca\textsuperscript{+}), and find evidence for hydrogen (H), sodium (Na), magnesium (Mg), neutral calcium (Ca), titanium (Ti), ionised titanium (Ti\textsuperscript{+}), ionised iron (Fe\textsuperscript{+}), neutral iron (Fe), and ionised strontium (Sr\textsuperscript{+}). The strongest detection was achieved in the red arm of Veloce, consistent with expectations due to the prominent Ca\textsuperscript{+} triplet at wavelengths around 850–870\,nm. Our results validate Veloce’s capability for high-resolution atmospheric studies, highlighting it as an accessible, flexible facility to complement larger international telescopes. If future observations stack multiple transits, Veloce has the potential to reveal atmospheric variability, phase-dependent spectral changes, and detailed chemical compositions of highly irradiated exoplanets.
\end{abstract}

\section{INTRODUCTION}
\label{sec:int}
High-resolution transmission spectroscopy using the cross-correlation technique~\citep{Snellen_2010_CC} has become a cornerstone method in exoplanet atmospheric studies. Since its initial demonstration, cross-correlation has enabled robust detections of atmospheric features by coherently combining numerous spectral lines, significantly increasing sensitivity to faint signals. The broad adoption of this method is evident from its successful application across diverse high-resolution spectrographs and telescopes, including HARPS-N at the TNG~\citep{Hoeijmakers_2018}, TRES at the Whipple Telescope~\citep{Lowson_2023}, ESPRESSO~\citep{Ehrenreich_2020} and UVES~\citep{Khalafinejad_2017_UVES_Detection} at the VLT, MAROON-X at Gemini-North~\citep{Pelletier_2023}, FOCES at Wendelstein~\citep{Borsato_2024}, FIES at the Nordic Optical Telescope~\citep{Bello_Arufe_2022}, and HARPS at La Silla~\citep{Wyttenbach_2015_HAPRS_detection}. Given this proven track record across multiple international facilities, there is clear motivation to test whether the Australian astronomical community’s Veloce spectrograph can similarly contribute to this growing field.

Much of the success of these observations stems from the discovery of ultra-hot Jupiters—an extreme subclass of an already extreme family of planets, with dayside temperatures exceeding 2200\,K~\citep{Parmentier_2018}. At such temperatures, their atmospheres undergo significant molecular dissociation, producing atomic gas-dominated transmission spectra~\citep{Heng_2017}. This results in strong optical absorption features from neutral and ionised species, making them particularly favourable targets for ground-based transmission spectroscopy.

WASP-189b exemplifies the ultra-hot Jupiter class, sitting near the extreme edge of their parameter space with an equilibrium temperature of 3353\,K and an orbital period of 2.72 days~\citep{Lendl_2020}. It orbits a bright A-type host star ($T_{\mathrm{eff}} = 7996\,\mathrm{K}$; $V = 6.6$; \citealt{Anderson_2018}) at a declination of $-3^\circ$, making it exceptionally observable from both hemispheres. The high stellar brightness combined with the planet’s intense irradiation produces strong, clearly resolved atmospheric signals, as demonstrated by numerous confirmed detections of atomic and ionised species~\citep{Yan_2020,Prinoth_2022,Stangret_2022,Prinoth_2023,Prinoth_2024a_Spectral_Atlas}. These qualities make WASP-189b an ideal benchmark for testing observational strategies and instrumentation capabilities.

Veloce, a fibre-fed, ultra-stable R4 echelle spectrograph mounted on the 3.9\,m Anglo-Australian Telescope (AAT) \citep{Gilbert_2018,Taylor_2024}, offers a design resolution of $R \sim 75{,}000$ and wavelength coverage from 396–950\,nm (396–480\,nm Azzurro; 470–650\,nm Verde; 630–950\,nm Rosso). Given successful atmospheric detections with telescopes of comparable or smaller aperture~\citep[e.g.][]{Bello_Arufe_2022,Lowson_2023,Borsato_2024}, Veloce represents a promising but currently untested capability for the Australian astronomical community. Demonstrating that Veloce can resolve exoplanet atmospheres within a single transit would significantly expand its scientific utility, providing timely and direct access to high-resolution atmospheric characterisation.

In this Letter, we test whether ground-based high-resolution transmission spectroscopy of exoplanet atmospheres is achievable with Veloce, using observations of a single transit of the ultra-hot Jupiter WASP-189b as a proof-of-concept.

\section{Observations and Data}
\label{sec:observations}
We observed a single transit of the ultra-hot Jupiter WASP-189b on 2024-04-10. Baseline spectra of the host star were acquired pre-ingress and post-egress, ensuring accurate normalisation of the stellar spectrum. During transit, airmass decreased from 2.0 to 1.1 at mid-transit, with stable wind conditions ($\sim$3m/s) enhancing throughput and instrumental stability. At the time, Veloce was limited to 2-amplifier readout mode due to a hardware issue, doubling readout overheads (120s per exposure). Exposure times were thus set at 500s, balancing photon collection with minimal radial-velocity smearing~\citep{Boldt_Christmas_2024}. We obtained 22 in-transit and 12 out-of-transit exposures, achieving median photon-limited S/N per pixel of 11.0 (Azzurro), 137.0 (Verde), and 204.0 (Rosso), with stable performance throughout. The laser frequency comb was not employed, as precise radial velocities were unnecessary.

\subsection{Data Reduction of the Raw Spectra}
\label{sec:data_reduction}
Raw spectra were reduced using the custom pipeline of \citet{Kruger_2025}, which applies minimal preprocessing (i.e. no barycentric correction, normalisation or order merging). Bias was removed from each quadrant via overscan regions, and pixel-to-pixel sensitivity variations were corrected using a normalised master flat constructed from all fibre flats taken that night (0.1\,s for Rosso, 1\,s for Verde and 60\,s for Azzurro). Scattered light subtraction was omitted, as it was negligible for our target. Spectral extraction proceeded by summing flux in the cross-dispersion direction around predefined traces modelled as fifth-order polynomials. Initial trace positions and dispersion-range definitions were taken from the Th traces in the Veloce Manual~\citep{VeloceManual}. To mitigate edge effects, the summing aperture was extended by three pixels beyond the flux drop-off observed in the fibre flats.

Wavelength calibration—common to all exposures—was derived from ThXe lamp spectra extracted at the start of the night. Air wavelengths of Th lines from the NIST database \citep{NIST_ASD} served as the absolute reference. An initial guess for line positions employed the converted-to-air solution provided by Chris Tinney in the Veloce Manual~\citep{VeloceManual}. Pixel positions of Th lines were then determined by fitting generalised Gaussians to features nearest these guesses above a relative intensity threshold. We fitted a seventh-degree polynomial surface to a data cube comprising pixel positions, absolute order numbers and air wavelengths scaled by order number, with iterative outlier rejection via 3\,$\sigma$ clipping of residuals. This solution was applied across the pixel range of each extracted order, yielding wavelength-calibrated spectra ready for CCF analysis. Figure~\ref{fig:reduced_spectrum} presents the resulting spectra for all three arms, plotted as wavelength against the square root of flux to emphasise variations in signal strength across the bandpass.

\subsection{1D Order Stitching}
We first cleaned each extracted 2D spectral order using a two-pass, sliding-window $\sigma$–clip: a 51-pixel median filter was applied to estimate the local continuum, flagging and replacing pixels exceeding $3\,\sigma$ from this baseline with the median value. To maintain high signal-to-noise, we retained only the top 80 per cent of flux values in each order. The cleaned orders were then continuum-normalised by dividing by a Savitzky–Golay-filtered continuum (window length 601 pixels, 3rd-order polynomial). Each pixel was assigned an inverse-variance weight based on the local flux standard deviation (computed within a 51-pixel window). All orders were subsequently interpolated onto a common wavelength grid spanning the full observed wavelength range, with overlapping regions combined via inverse-variance weighting. Remaining outliers were identified and removed using Astropy’s \texttt{sigma\_clip}~\citep{Astropyy_2022}. Finally, the stitched 1D spectrum was smoothed by convolution with a Gaussian kernel matched to the spectrograph’s resolution ($R = 75,000$).

\subsection{Data Post-processing}\label{sect:data_processing}
To correct for telluric absorption, we used the radiative transfer code \texttt{Molecfit}~\citep{Smette_2015,Kausch_2015} to model and remove atmospheric features from each exposure individually. We aligned the one-dimensional spectrum using the telluric lines before performing a chi-squared fit to match the tellurics present in the data. A synthetic telluric model was then generated and divided out from each individual exposure, removing the imprint of Earth’s atmosphere. After telluric correction, we compute an out-of-transit master spectrum and divide each exposure by this baseline to isolate the planetary signal, producing a time series of residual spectra. These residuals are primarily shaped by variations in continuum levels between exposures, resulting in non-flat baselines. We flatten each spectrum using a high-pass filter with a width of 80~km\,s$^{-1}$. A running median absolute deviation (MAD) filter with a width of 20 pixels is then applied: pixels deviating by more than 5$\sigma$ from the local median are flagged as outliers and replaced via linear interpolation. Finally, the spectra are visually inspected for remaining contaminants, such as interstellar absorption features, which are manually masked with \texttt{NaNs}. This observational setup follows the procedure outlined in \citet{Borsato_2023}.

\section{METHODS: The Cross-Correlation Technique}
With the residual spectra prepared in Section~\ref{sect:data_processing}, we employ the cross-correlation technique~\citep{Snellen_2010_CC} to search for the planetary signal embedded within the noise. The process involves computing a series of weighted averages of the residual spectra, where the weights are given by the predicted absorption profile of a target species, such as an atomic or ionised metal. These spectrum-weighted averages are evaluated across a grid of Doppler velocities; when the planet’s radial velocity aligns with a particular Doppler shift, the weighted average is maximised, revealing the presence of the species in the planet’s atmosphere.

Mathematically, it is defined as;

\begin{equation}
    c(v) = \sum_{i=0}^{N} x_i \hat{T}_i(v)
\label{eq:ccf}
\end{equation}

where $c(v)$ is the weighted average of the flux $x_i$, and $\hat{T}_i(v)$ is the template spectrum providing the velocity-dependent weights in the summation.

\begin{equation}
    \hat{T}_i(v) = \frac{T_i(v)}{\Sigma T_i(v)}
\end{equation}

where $\hat{T}_i(v)$ is the normalised template spectrum, $T_i(v)$ represents the original template values, and $\sum T_i(v)$ is the total flux across the template. The MANTIS templates~\citep{Kitzmann_2021} are a library of cross-correlation templates for numerous atomic and ionised species, spanning 320--5500\,nm. We use these templates for our analysis, selecting those generated at an equilibrium temperature of 4000\,K. In this study, we cross-correlate with all species detected in~\cite{Prinoth_2023} who reported H, Na, Mg, Ca, Ca\textsuperscript{+}, Ti, Ti\textsuperscript{+}, TiO, V, Cr, Mn, Fe, Fe\textsuperscript{+}, Ni, Sr, Sr\textsuperscript{+}, and Ba\textsuperscript{+} in the planet's atmosphere.

\begin{figure}[t]
    \centering
    \includegraphics[width=1.0\textwidth]{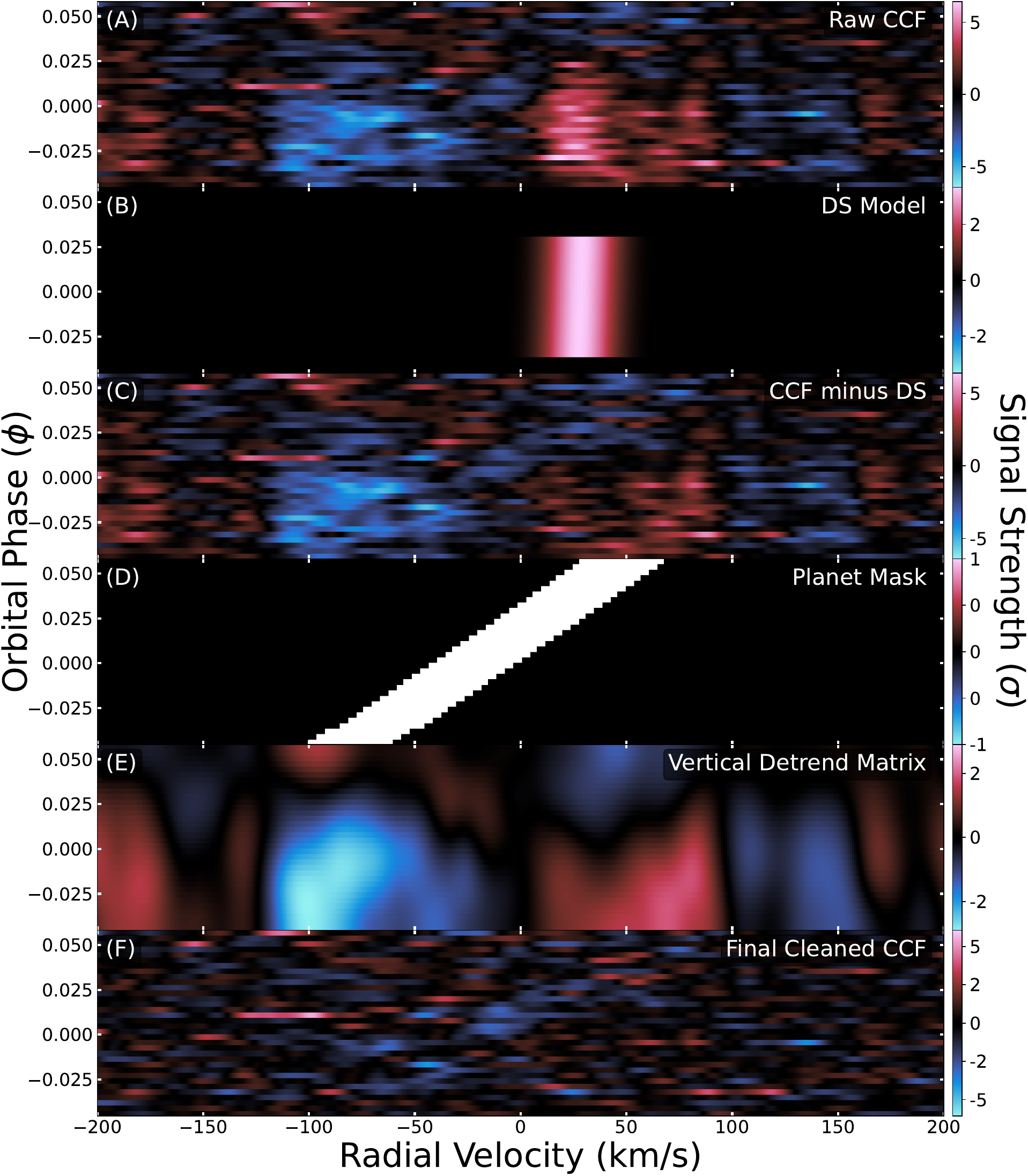}
    \caption{Step-by-step illustration of the cross-correlation function (CCF) cleaning pipeline applied to one night of WASP-189b observations using the combined template. \textbf{(A)} Raw CCF map showing the bright, time-varying Rossiter–McLaughlin (RM) effect. \textbf{(B)} Best-fit Gaussian model of the Rossiter–McLaughlin (RM) effect. \textbf{(C)} CCFs after subtracting the RM model. \textbf{(D)} Binary mask used to exclude the planetary signal from detrending. \textbf{(E)} Fifth-degree polynomial detrending surface fitted along the phase axis, masking the planetary trail. \textbf{(F)} Final cleaned CCF map. Colour represents signal strength in units of the continuum standard deviation, with blue indicating absorption and green indicating emission.}
    \label{fig:cleaning_steps_appled}
\end{figure}

\subsection{CCF Cleaning and Detrending}
Due to the sensitivity of cross-correlation, the process can amplify small residual artefacts in the data. Two dominant signals often appear: line distortions in the stellar spectrum caused by the Rossiter–McLaughlin (RM) effect, and residuals from imperfect telluric correction by \texttt{Molecfit}. To address these, we apply additional standard cleaning procedures~\citep{Hoeijmakers_2019,Prinoth_2022,Borsato_2023,Prinoth_2023}. The RM effect appears in the CCF as a bright, time-dependent Gaussian feature. We remove this by forward-modelling a time-varying Gaussian using an MCMC sampler with informed priors on its amplitude and velocity offset. The best-fit Gaussian is then subtracted from the raw CCF map, effectively eliminating the RM signal. In principle, this feature can be modelled directly in the residual CCFs (as in tomographic studies such as \citealt{Rainer_2021}) to extract the same parameters usually derived from RM analyses, such as the projected obliquity and $v \sin i_\star$. In the present work we remove it in order to recover a coherent planetary signal in the velocity–velocity diagrams. In future work, we plan to carry out a dedicated analysis of this dataset to investigate the RM effect in detail, alongside other observables that we speculate may be accessible (see Section \ref{sec:discussion}).

To mitigate telluric residuals, we fit a fifth-degree polynomial along the orbital phase axis at each velocity step. This removes vertical patterns in the CCF. Prior to fitting, we mask the expected planetary trail using \texttt{NaNs} to ensure the fit is not biased. Subtracting this two-dimensional trend surface yields a fully cleaned CCF map. The full cleaning procedure is illustrated in Figure~\ref{fig:cleaning_steps_appled}.

\subsubsection{Constructing Velocity–Velocity Diagrams}
Velocity–velocity diagrams (often referred to as $K_p$–$V_{\text{sys}}$ maps) enhance the detection fidelity of the cross-correlation function~\citep{Brogi_2012}. They represent co-added cross-correlation exposures, combining all observations into a single detection signal. Assuming a static and symmetrical atmosphere surrounding the planet, its radial velocity can be computed as:

\begin{equation}
v_r(\phi) = V_{\mathrm{rv}} + V_{\mathrm{grad}} \,\sin i \,\sin\bigl(2\pi\phi\bigr)
\;;\quad
\phi = \frac{t - T_0}{P}
\label{eq:vv_diagram}
\end{equation}

where $v_r(\phi)$ is the planet’s radial velocity at orbital phase $\phi$, $t$ is the time stamp of observation, $T_0$ is the mid-transit time, and $P$ is the orbital period. $V_{\mathrm{rv}}$ is a radial velocity offset, $V_{\mathrm{grad}}$ is a time-dependent velocity gradient, and $i$ is the orbital inclination. By applying Equation~\ref{eq:vv_diagram} across a grid of trial $V_{\mathrm{grad}}$ values, we shift each CCF exposure into hypothetical planetary rest frames and average them. The resulting 2D map provides a clear visual confirmation of any atmospheric signal (see Figure~\ref{fig:CaII_detection}). Note that $V_{\mathrm{rv}}$ and $V_{\mathrm{grad}}$ often correspond to the systemic velocity ($V_{\mathrm{sys}}$) and orbital semi-amplitude ($K_p$), respectively, but can differ due to atmospheric geometry and dynamical effects.

\subsection{Model Injection}
To evaluate whether the absence of a signal reflects a true non-detection or stems from limitations in our analysis, we perform model injection. This technique has become standard in high-resolution exoplanet spectroscopy \citep{Snellen_2010_CC,Birkby_2013,Hoeijmakers_2018_tau_bootis}.

For the injection–recovery tests, each of the 17 MANTIS templates at 4000 K was injected individually into the observed spectra prior to CCF computation. This allows us to evaluate, species by species, whether a signal matching that template would be detectable given our S/N and observing conditions. Injected models were centred on the literature values of the systemic velocity ($V_{\rm sys} = -26.8$ km\,s$^{-1}$) and orbital semi-amplitude ($K_p = 202$ km\,s$^{-1}$; \citealt{Anderson_2018}).

Because the MANTIS templates assume a fiducial hot-Jupiter system ($R_p = 1.5\,R_J$, $R_\star = 1\,R_\odot$, \citealt{Kitzmann_2021}), their absolute line depths must be rescaled to represent WASP-189b. We therefore apply a correction factor based on the squared radius ratio,

\begin{equation}
    f = \left(\frac{(R_p/R_\star)_{\mathrm{WASP-189A}}}{(R_p/R\star)_{\mathrm{MANTIS}}}\right)^2 .
\end{equation}

All templates are scaled by this factor before injection to ensure their line depths are consistent with WASP-189b’s true geometry. For WASP-189b ($R_p = 1.619\,R_J$, $R_\star = 2.362\,R_\odot$, \citealt{Lendl_2020}), $f \approx 0.219.$

Injections allow us to verify whether our data quality and channel throughput are sufficient to recover the expected signal. In contrast, for newly discovered planets or species with no prior detections, theoretical templates generated via radiative transfer models can be injected to determine whether a non-detection reflects true atmospheric absence or insufficient sensitivity.

These models are broadened to account for instrumental resolution and planetary rotation, and resampled to match the Veloce wavelength grid. We then reprocess the injected spectra through our full pipeline, including telluric correction, continuum normalisation, and cross-correlation. By comparing the injected results to our original data, we assess whether a species should have been detectable given our signal-to-noise and observing conditions. This allows us to distinguish between non-detections driven by physical absence and those due to insufficient sensitivity or loss of signal from systematics.

\subsection{Combining the Channels of Veloce}
\label{sec:combine_arms}
For each spectrograph arm (Azzurro, Verde and Rosso), we begin by loading the cross-correlation function (CCF) map. We measure the noise level by computing the standard deviation across the edges of the velocity grid, where no planetary signal is expected. We identify a small region in radial velocity–velocity gradient space around the expected planetary signal. Within this region, we measure the average strength of the normalised signal and, where available, compare it to the injected-model prediction. These values are used to assign a raw weight to each arm. Negative weights are discarded, and the remaining weights are normalised so that their sum equals one. This ensures that the contribution from each spectrograph arm is proportional to its signal content.

\section{RESULTS}
\begin{figure}[htbp]
    \centering
    \includegraphics[width=1.0\textwidth]{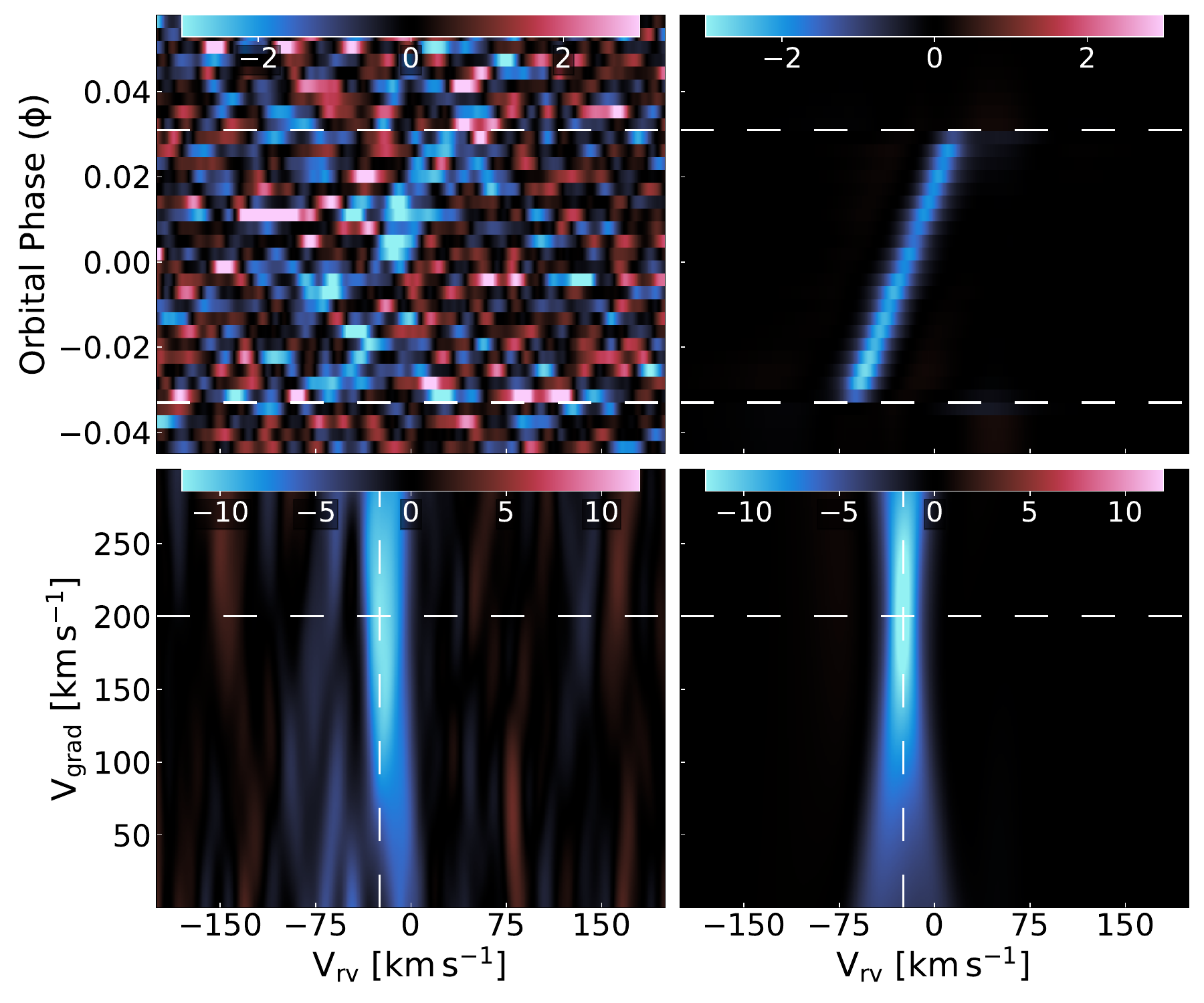}
    \caption{
        Detection of ionised calcium (Ca\textsuperscript{+}) in the atmosphere of WASP‑189b.
        \textbf{Top row:} 2D cross‐correlation function (CCF) maps for the data (left) and the expected signal from model injection (right).
        \textbf{Bottom row:} Corresponding velocity-velocity diagrams (data left; expected signal right).
        Dashed lines indicate the systemic velocity ($V_{\rm sys} = -26.8\,$km\,s$^{-1}$) and semi‐amplitude ($K_p = 202\,$km\,s$^{-1}$)}.
    \label{fig:CaII_detection}
\end{figure}

\begin{figure*}[htbp]
    \centering
    \includegraphics[width=1.0\textwidth]{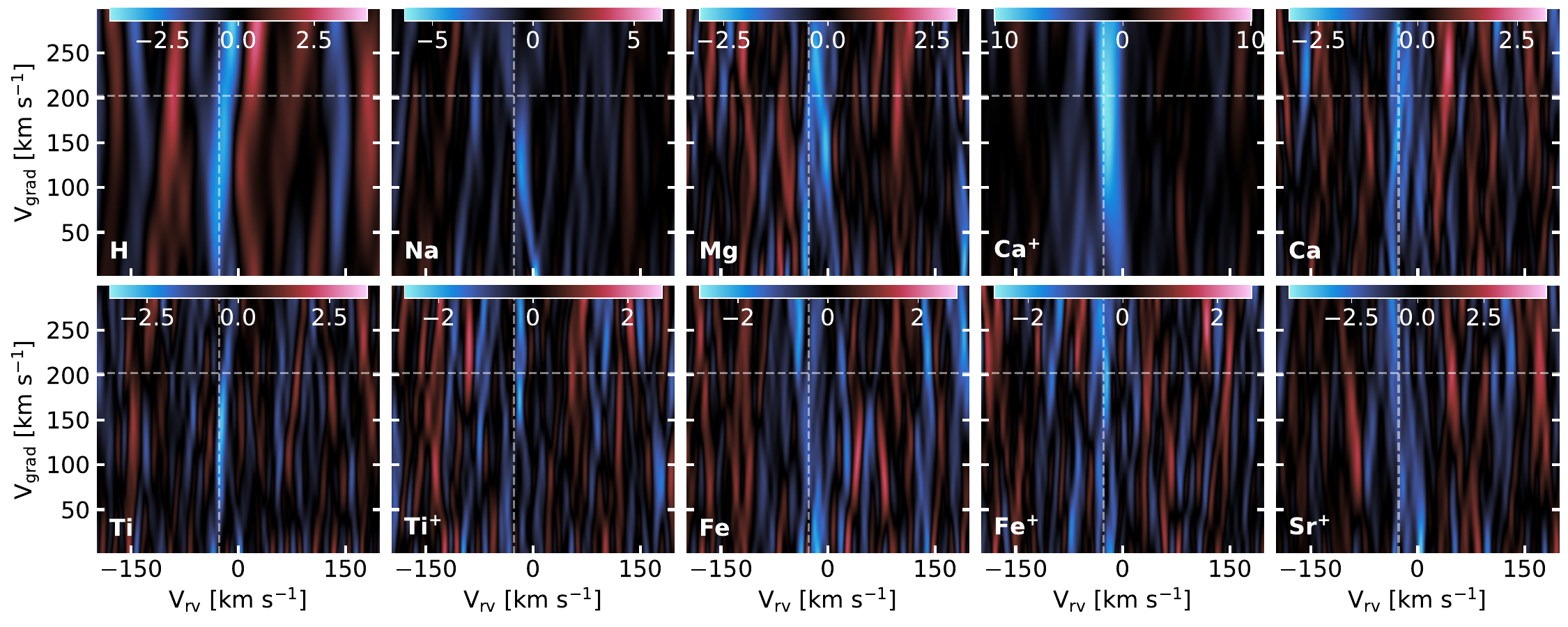}
    \caption{Mosaic of velocity-velocity diagrams for the species robustly detected in a single transit of WASP-189b with Veloce. Each panel shows the combined CCF signal for H, Na, Mg, Ca\textsuperscript{+}, Ca, Ti, Ti\textsuperscript{+}, Fe\textsuperscript{+}, Fe and Sr\textsuperscript{+}, arranged by increasing atomic number from top left to bottom right. White dashed lines mark the expected systemic velocity ($V_{\rm sys}=-26.8\,$km\,s$^{-1}$) and orbital semi‐amplitude ($K_p=202\,$km\,s$^{-1}$). The colour scale indicates cross‐correlation strength in units of the continuum standard deviation.}
    \label{fig:detection}
\end{figure*}

\subsection[The detection of Ca+]{The detection of Ca\textsuperscript{+}}
We detect a clear atmospheric signal of ionised calcium (Ca\textsuperscript{+}) in WASP-189b, partially visible in the cross-correlation function (CCF) and robustly in the velocity–velocity diagram shown in Figure~\ref{fig:CaII_detection}. The planetary trace appears slightly redshifted in $V_{\rm rv}$ compared to the expected systemic velocity, while the peak $V_{\rm grad}$ is offset to somewhat lower values than the expected orbital semi-amplitude. The signal in the velocity-velocity diagram exceeds $5\sigma$ in amplitude, based on the standard deviation of background regions away from the planetary trail. The injected model predicts a comparative detection significance as well, suggesting that with the current throughput of Veloce we would expect the signal to be this strong. Among known species in this planet’s atmosphere, Ca\textsuperscript{+} is the strongest contributor to optical absorption~\citep{Prinoth_2023}, and our result represents a robust agreement to this fact using a single transit with Veloce.

\subsection{Additional Species Detections}
Beyond Ca\textsuperscript{+}, we also produce evidence for absorption at the expected velocities for H, Na, Mg, Ti, Ti\textsuperscript{+} and Fe\textsuperscript{+} (Figure~\ref{fig:detection}). These panels exhibit coherent absorption peaks in the cross‐correlation maps, indicative of possible planetary absorption. Other tested species—Ca, Fe and Sr\textsuperscript{+}—show broad depressions at the correct velocities but lack distinct signatures. While these signals warrant further investigation, robust confirmation will require stacking multiple transits to enhance signal‐to‐noise. Many of these signals have peaks at lower values than the expected $K_p$ of WASP-189b. This is a common outcome in cross-correlation studies and is attributed to the asymmetric geometry of the planet’s inflated, highly irradiated dayside~\citep[e.g.][]{Prinoth_2022}.

We compare the relative performance of Veloce’s three arms using the Fe\textsuperscript{+} signal as a reference (see Figure~\ref{fig:comparison_across_arms}). The green (Verde) arm contributes the strongest Fe\textsuperscript{+} absorption, and the blue (Azzurro) arm also shows a significant contribution, reflecting deeper Fe\textsuperscript{+} line strengths in covered wavelength region. In contrast, the red (Rosso) arm contributes the least, as the Fe\textsuperscript{+} lines it covers are relatively shallow, despite the higher signal fidelity. This comparison highlights the importance of tailoring future observational strategies to optimise sensitivity to ionised metals.

For completeness and transparency, we present the mosaic of non‐detections in Figure~\ref{fig:non_detections}. These panels demonstrate the absence of clear signals for V, Cr, Mn, Ni and Ba\textsuperscript{+}, underscoring the need for higher S/N or multiple‐transit datasets to explore these species further. While TiO appears to show a feature, we do not interpret this as a detection. Our injected models do not predict a measurable TiO signal, and prior studies~\citep[e.g.][]{Prinoth_2022} have shown that robust TiO detections typically require many stacked transits.

\section{DISCUSSION \& CONCLUSION}
\label{sec:discussion}
In this study, we present the first detection of an exoplanet atmosphere using the Veloce spectrograph, marking a significant milestone for high-resolution spectroscopy with Australian instrumentation. Our detection of ionised calcium (Ca\textsuperscript{+}), along with evidence for absorption by H, Na, Mg, Ti, Ti\textsuperscript{+} and Fe\textsuperscript{+}, aligns with species previously studied in WASP-189b’s atmosphere~\citep{Prinoth_2023}. These detections validate our cross-correlation pipeline and confirm Veloce’s potential for atmospheric characterisation.

Our detection of Ca\textsuperscript{+} matches expectations from prior studies~\citep{Prinoth_2023}, owing to its strong optical triplet in the Rosso channel. The observed detection significance of $>5\sigma$ reflects the empirical signal-to-noise ratio of the Ca\textsuperscript{+} trail relative to the background scatter in the processed CCF. The injected model predicts a comparable detection significance, showing that the observed signal strength is in line with expectations given Veloce’s current throughput, and the agreement indicates that our detections are consistent with both the data quality and prior studies (e.g. \citealt{Prinoth_2023}, who reported $>25\sigma$ for Ca\textsuperscript{+} using higher signal-to-noise datasets). Our injection–recovery tests predict $\lesssim 1\sigma$ peaks for the non-detections (e.g. V, TiO, Sr\textsuperscript{+}, Ni, Mn) in a single transit, below the noise floor; the null results are therefore sensitivity-limited rather than physically constraining, and do not rule out reasonable abundances.

We find that the green (Verde) and red (Rosso) arms deliver comparable absorption signals for species with strong line coverage in their respective wavelength ranges, while the blue (Azzurro) arm shows weaker performance due to its lower throughput. Despite this, Azzurro still contributes measurable signal, underscoring the species-dependent nature of arm sensitivity. These results suggest that optimising target selection and fibre alignment for each channel will maximise Veloce’s overall atmospheric sensitivity.

Veloce combines the advantages of fibre-fed stability and the 3.9-m aperture of the AAT, enabling frequent observational opportunities without the scheduling pressures of larger international facilities. With approximately ten observable transits of WASP-189b each year, it becomes feasible to stack multiple transits, significantly enhancing detection sensitivity and enabling robust chemical inventories of exoplanet atmospheres~\citep[e.g.][]{Borsato_2023}. Additionally, repeated observations across multiple epochs would allow studies of atmospheric variability and phase-dependent line variations, as demonstrated in recent work by~\cite{Prinoth_2023, Prinoth_2024a_Spectral_Atlas}.

Future campaigns should focus on optimising fibre alignment—especially in the blue channel—to boost throughput and sensitivity to short‑wavelength metal lines. Even modest gains in instrumental efficiency could enable abundance determinations, atmospheric‑structure studies and time‑resolved single‑line spectroscopy~\citep[e.g.][]{Prinoth_2024a_Spectral_Atlas,Vaulato_2025}, demonstrating that Veloce, though designed for precision radial‑velocity work, offers compelling capabilities for exoplanet atmospheric characterisation.

\begin{acknowledgement}
Based on observations collected at the Anglo-Australian Telescope under program \texttt{A/2024B/003}. We acknowledge the traditional custodians of the land on which the AAT stands, the Gamilaraay people, and pay our respects to their Elders past, present and emerging. Nicholas W. Borsato led the data analysis, manuscript writing, and typesetting. Joachim Kr\"uger reduced the raw Veloce spectra and contributed to the data reduction section. Nicholas W. Borsato, Daniel B. Zucker, and Sarah L. Martell conducted the observations. Simon J. Murphy and Duncan Wright provided valuable feedback and suggestions that improved the manuscript.

JK was supported by the UniSQ International Stipend and UniSQ International Fees Research Scholarship. SJM was supported by the Australian Research Council through Future Fellowship FT210100485. SLM acknowledges support from ARC DP220102254 and the UNSW Scientia Fellowship programme.
\end{acknowledgement}


\bibliography{bibtemplate}

\appendix
\renewcommand{\thefigure}{A\arabic{figure}}
\setcounter{figure}{0}

\section{Additional Data Processing Plots}
\begin{figure*}[htbp]
    \centering
    \includegraphics[width=0.9\textwidth]{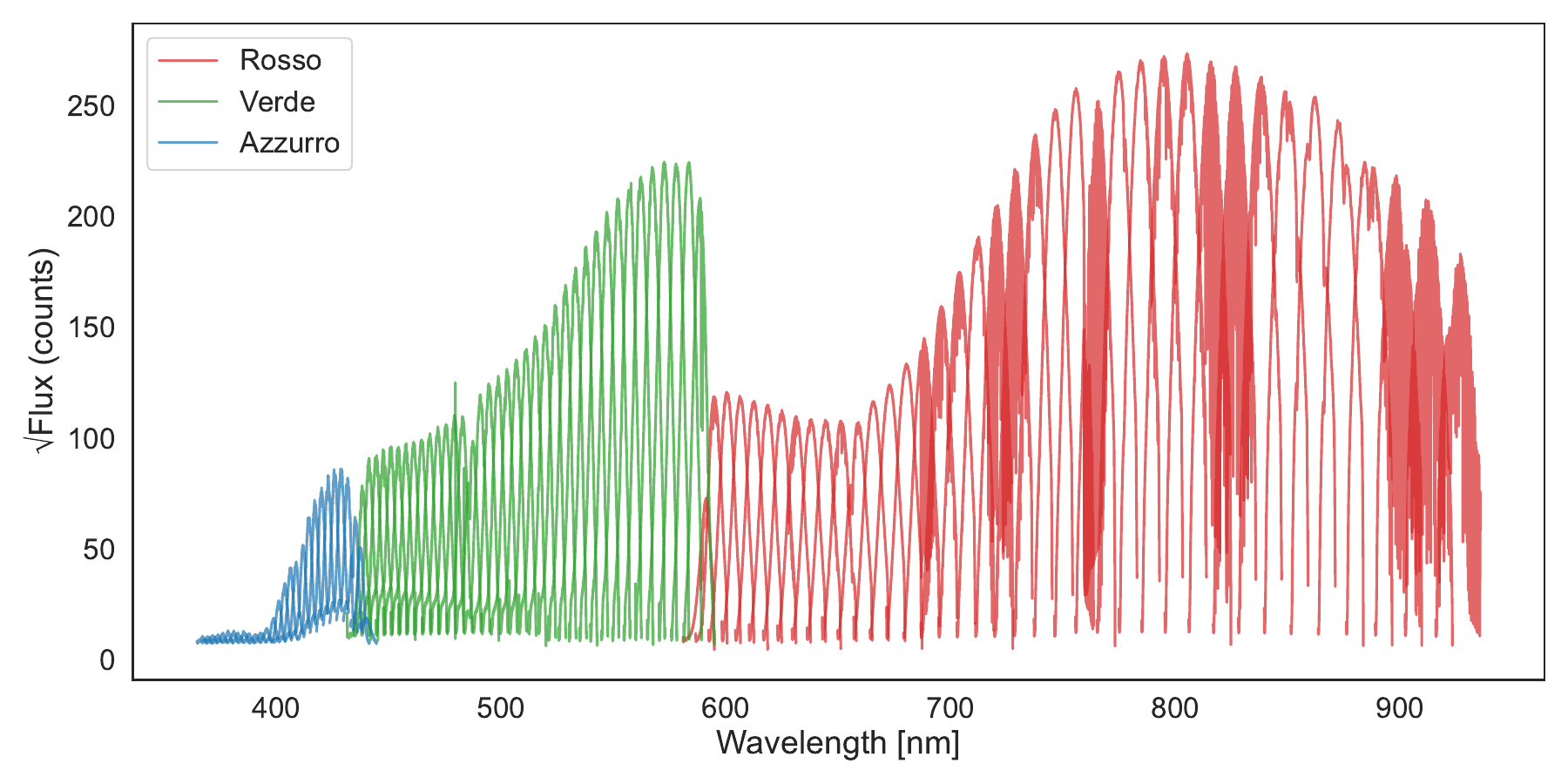}
    \caption{Raw 2D echelle spectra from the Veloce spectrograph showing extracted blaze-function profiles for each order. These spectra represent pipeline-extracted counts prior to continuum normalisation and order stitching, illustrating the raw blaze shape and signal distribution across the detector.}
    \label{fig:reduced_spectrum}
\end{figure*}

To provide context for our reduction workflow, Figure~\ref{fig:reduced_spectrum} shows the raw 2D echelle spectrum produced by the Veloce pipeline. The blaze-function shape and signal distribution across the detector are visible prior to continuum normalisation and stitching. Additionally, Figure~\ref{fig:cleaning_steps_appled} provides a visual summary of the cross-correlation function (CCF) cleaning steps applied to remove stellar and telluric residuals from the processed spectra.

\section{CCF Maps of Alternate Species}
\label{sec:alt_species}

\begin{figure*}[htbp]
    \centering
    \includegraphics[width=1.0\textwidth]{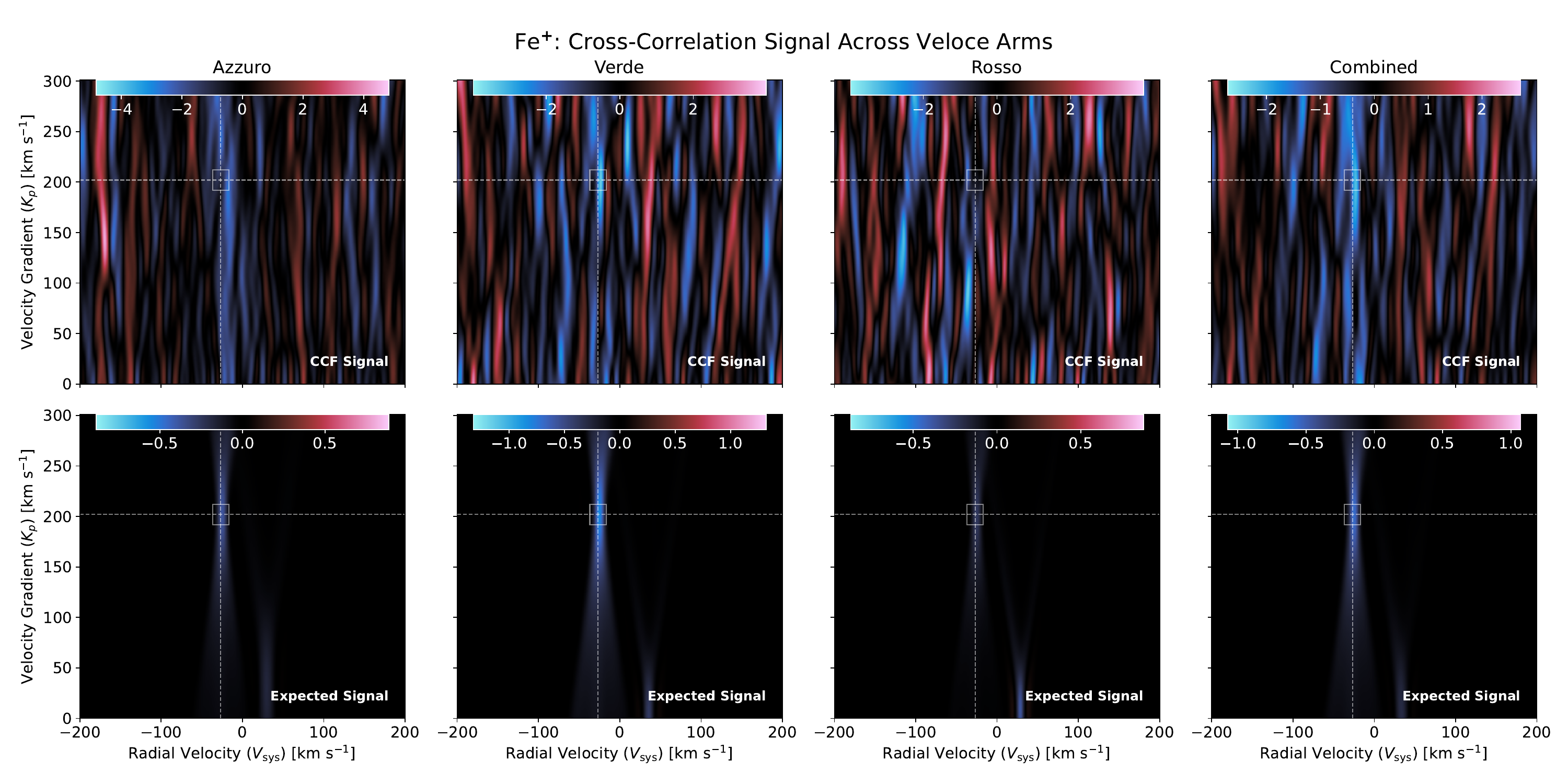}
    \caption{
    Velocity–velocity maps for Fe cross-correlation detections across the blue, green, and red arms of the Veloce spectrograph. Dashed lines indicate the expected systemic velocity ($V_{\rm sys} = -26.8\,$km\,s$^{-1}$) and semi-amplitude ($K_p = 202\,$km\,s$^{-1}$).
    }
    \label{fig:comparison_across_arms}
\end{figure*}

\begin{figure*}[htbp]
    \centering
    \includegraphics[width=1.0\textwidth]{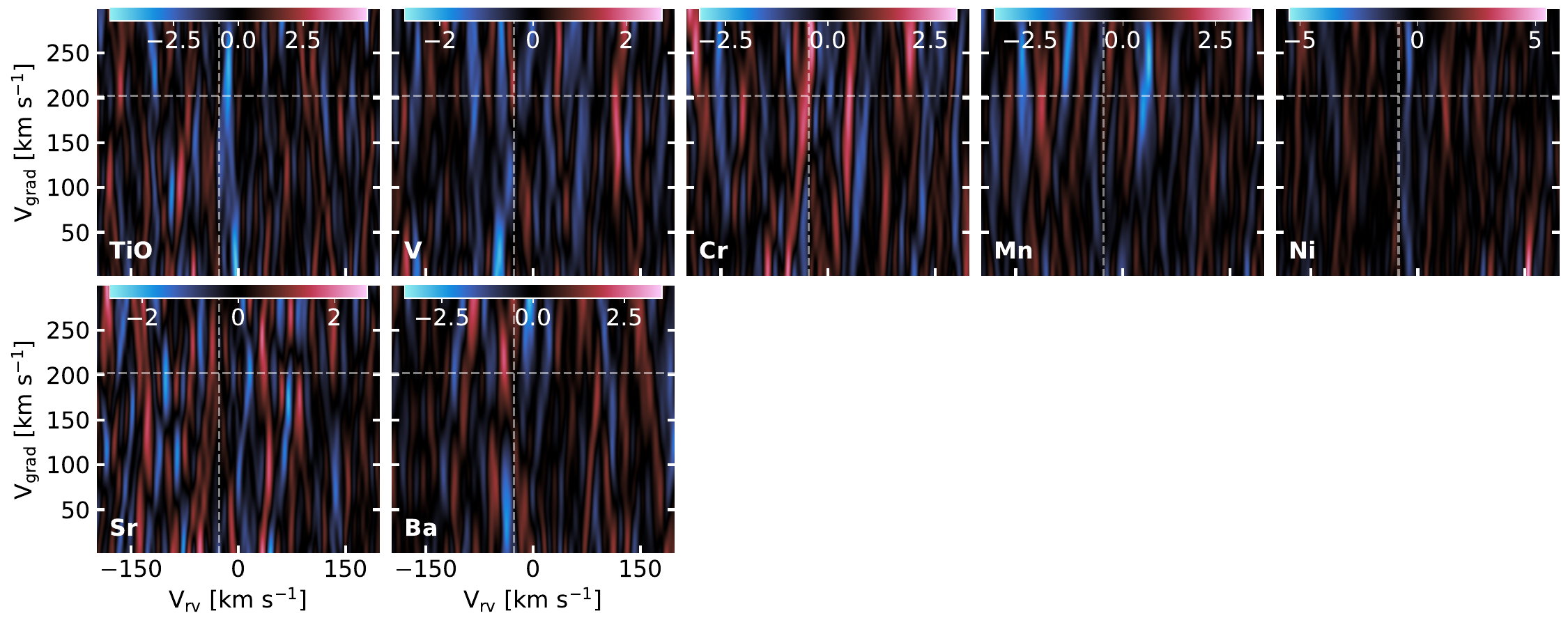}
    \caption{Mosaic of velocity-velocity diagrams for species with no secure atmospheric detection in a single transit of WASP-189b using Veloce. Each panel shows the combined CCF signal for TiO, V, Cr, Mn, Ni, Sr and Ba\textsuperscript{+}, arranged by increasing atomic number from top left to bottom right. White dashed lines mark the expected systemic velocity ($V_{\rm sys}=-26.8\,$km\,s$^{-1}$) and orbital semi‐amplitude ($K_p=202\,$km\,s$^{-1}$). The colour scale indicates cross‐correlation strength in units of the continuum standard deviation.}
    \label{fig:non_detections}
\end{figure*}

 Beyond our primary detection of Ca\textsuperscript{+}, we examined the velocity-velocity diagrams for other species to evaluate marginal and non-detections. Figure~\ref{fig:comparison_across_arms} shows the performance across spectrograph arms using Fe\textsuperscript{+} as a representative species. Meanwhile, Figure~\ref{fig:non_detections} displays a mosaic of velocity-velocity diagrams for species without clear detections in our single-transit analysis, including TiO, V, Cr, Mn, Ni, Sr, and Ba\textsuperscript{+}. These figures illustrate both the variability in arm performance and the limitations of single-transit sensitivity for weaker species. Additional figures showing individual maps are provided in the appendix.

\end{document}